\documentclass[twocolumn,showpacs,superscriptaddress,aps,prb]{revtex4-1}
\usepackage{amsfonts}
\usepackage{graphicx}
\usepackage{times}
\usepackage{amssymb}
\usepackage{amsmath}

\begin{document}

\title{Emergent infinite-randomness fixed points from the extensive random
bipartitions of the spin-1 Affleck-Kennedy-Lieb-Tasaki topological state }
\author{Min Lu}
\author{Wen-Jia Rao}
\affiliation{Zhejiang Institute of Modern Physics, Zhejiang University, Hangzhou 310027,
China}
\author{Rajesh Narayanan}
\affiliation{Department of Physics, Indian Institute of Technology Madras, Chennai
600036, India}
\affiliation{Asia Pacific Center for Theoretical Physics, Pohang, Gyeongbuk 790-784, Korea}
\author{Xin Wan}
\affiliation{Zhejiang Institute of Modern Physics, Zhejiang University, Hangzhou 310027,
China}
\affiliation{Collaborative Innovation Center of Advanced Microstructures, Nanjing 210093,
China}
\author{Guang-Ming Zhang}
\affiliation{State Key Laboratory of Low-Dimensional Quantum Physics and Department of
Physics, Tsinghua University, Beijing 100084, China}
\affiliation{Collaborative Innovation Center of Quantum Matter, Beijing, China}
\date{\today }

\begin{abstract}
Quantum entanglement under an extensive bipartition can reveal the critical
boundary theory of a topological phase in the parameter space. In this study
we demonstrate that the infinite-randomness fixed point for spin-1/2 degrees
of freedom can emerge from an extensive random bipartition of the spin-1
Affleck-Kennedy-Lieb-Tasaki chain. The nested entanglement entropy of the
ground state of the reduced density matrix exhibits a logarithmic scaling
with an effective central charge $\tilde{c} = 0.72 \pm 0.02 \approx \ln 2$.
We further discuss, in the language of bulk quantum entanglement, how to
understand all phase boundaries and the surrounding Griffiths phases for the
antiferromagnetic Heisenberg spin-1 chain with quenched disorder and
dimerization.
\end{abstract}

\maketitle

\section{INTRODUCTION}

Entanglement spectrum under an extensive bipartition of a topological ground
state has recently emerged as a novel approach to study the quantum phase
transition between the topological phase and its trivial counterpart.~\cite%
{hsieh13,rao14,borchmann14,hsieh14,santos14,zhu14,vijay14,rao16} The so-called
bulk entanglement spectrum (BES) reveals the boundary theory in the
corresponding parameter space of a model system, rather than the edge states
along the fictitious boundary in real space.~\cite{Li-Haldane,Qi-2012} These
studies suggest that a triangular correspondence among the bulk theory, the
edge theory, and the critical theory may exist generically for a topological
phase.~\cite{Chen-Wang-Lu-Lee,zhu14}

\begin{figure}[t]
\includegraphics[width=7.5cm]{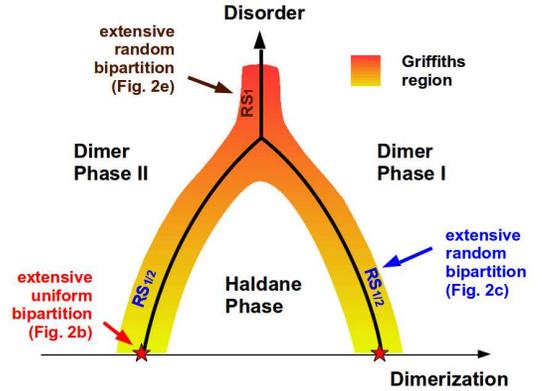}
\caption{\label{fig:phasediagram} (Color online) Phase diagram of the
  antiferromagnetic Heisenberg spin-1 chain in the presence of
  quenched bond disorder and dimerization.~\cite{damle02} The regimes
  of interest for the extensive bipartitions in
  Fig.~\ref{fig:partition} are labeled accordingly. The red stars mark
  the critical spin-1/2 chain that can be accessed by the extensive
  uniform bipartition of the AKLT state.~\cite{rao14} The thick black
  boundaries are the random singlet states that can be accessed by
  extensive random bipartitions of the AKLT state (see text). }
\end{figure}

An instructive example is the spin-1 Haldane gapped phase,~\cite%
{Haldane-1983} whose fixed-point properties are captured by the valence-bond
solid picture of the Affleck-Kennedy-Lieb-Tasaki (AKLT) model Hamiltonian.~%
\cite{AKLT} The corresponding exact ground state wave function can be
expressed as a matrix product state (MPS), indicating that the fundamental
entities of the spin-1 chain is actually the fractionalized spinons carrying
spin 1/2 that form singlets across adjacent sites. The spin-1/2 object can
be observed as the edge mode at the end of the chain, or at the end of a
partition in the entanglement study. The degeneracy due to the
fractionalized spin in the entanglement spectrum of the antiferromagnetic
Heisenberg spin-1 chain is key to understand the Haldane phase as a symmetry
protected topological phase.~\cite{Pollmann-Oshikawa-2010} Under a uniform
extensive bipartition with disjoint segments, the segment end spins coalesce
into an emergent critical spin-1/2 chain with a central charge $c = 1$,~\cite%
{rao14} which describes the collapse of the Haldane phase under imposed
dimerization\cite{rao16} (marked by the stars in Fig.\ref{fig:phasediagram}).

Meanwhile, spin chains can be interesting with a random probability
distribution of the bond couplings between neighboring spins, which may be
broaden without limit as the system is coarse-grained.~\cite{huse01} Such a
system is governed by the infinite-randomness fixed point (IRFP), whose
ground state is characterized by a random pattern of spin singlets formed
over large spatial separations.~\cite{dasgupta79,fisher94} For the
antiferromagnetic Heisenberg spin-1 chain the Haldane phase is stable
against weak disorder~\cite{hyman97,monthus97} and weak dimerization,~\cite%
{affleck87} but the emergent critical spin-1/2 chain is not.~\cite%
{dasgupta79,fisher94} As illustrated in Fig.~\ref{fig:phasediagram}, the
Haldane phase and two dimer phases in the presence of quenched disorder are
separated by the spin-1/2 random singlet (RS$_{1/2}$) boundaries, which
merges into a single spin-1 random singlet (RS$_1$) boundary between the
dimer phases. In addition, unusual Griffiths effects, characterized by two
continuously varying dynamical exponents, appear near the boundaries.~\cite%
{damle02}

In this paper we will show that disorder physics can also be revealed in the
bulk entanglement study under an extensive random bipartition of the spin-1
chain. For uncorrelated segment length, the nearest-neighbor couplings in
the bulk entanglement Hamiltonian exhibits a power-law distribution, which
is precisely the fixed-point solution for random antiferromagnetic spin-1/2
chains under strong-disorder renormalization-group (SDRG) transformation.~%
\cite{fisher94} Therefore, the ground state of the entanglement Hamiltonian
realizes the RS$_{1/2}$ state, for which we provide further evidence by
fitting the ensemble-averaged nested entanglement entropy to a logarithmic
scaling with an effective central charge $\tilde{c} = 0.72 \pm 0.02 \approx
\ln 2$. We discuss how to vary the random bipartition to explore the phase
diagram of the antiferromagnetic Heisenberg spin-1 chain with quenched
disorder and dimerization.

\section{The Affleck-Kennedy-Lieb-Tasaki spin-1 chain}

The spin-1 AKLT parent Hamiltonian on a periodic chain of length $L$ is
defined by~\cite{AKLT}
\begin{equation}  \label{eqn:aklt}
H_{\mathrm{AKLT}}=\sum_{i=1}^L J \left[ \mathbf{s}_{i} \cdot \mathbf{s}%
_{i+1}+ \frac{1}{3}\left( \mathbf{s}_{i} \cdot \mathbf{s}_{i+1}\right) ^{2}%
\right],
\end{equation}
whose exact ground state can be expressed as an MPS
\begin{equation*}
\left| \Psi_{\mathrm{AKLT}} \right\rangle =\sum_{\{s_{i}\}} \text{Tr}\left(
\mathbf{A }^{\left[ s_{1}\right] }\mathbf{A}^{\left[ s_{2}\right] } \cdots
\mathbf{A}^{ \left[ s_{L}\right] }\right) \left| \text{s}_{1} s_{2} \cdots
s_{L} \right\rangle ,
\end{equation*}
for $J>0$, where the local physical spin $s_{i}=-1,0,+1$ and $\mathbf{A}^{%
\left[ s \right] }$ are local $2 \times 2$ matrices given, e.g., in Ref.~%
\onlinecite{rao14}. In the thermodynamic limit, the spin-spin correlation
function decays exponentially with a correlation length $\xi = 1/\ln 3
\approx 0.91$, and any two spins that are separated by an even number of the
lattice sites are antiferromagnetically correlated.

\begin{figure}[t]
\includegraphics[width=7.5cm]{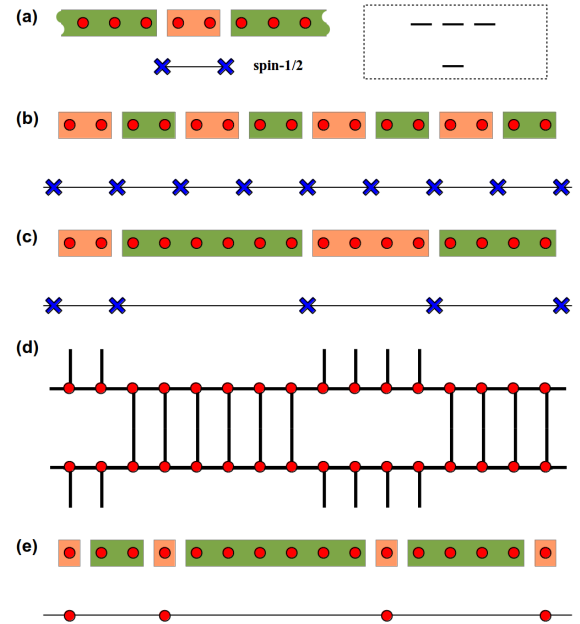}
\caption{\label{fig:partition} (Color online) Various bipartitions of
  the AKLT state and the resulting effective spin models for the
  entanglement Hamiltonian. We denote A the collection of segments
  with orange shadow and B that of the segments with green shadow. (a)
  The entanglement spectrum (illustrated in the box) of a finite
  segment with an even number of sites is equivalent to the spectrum
  of two antiferromagnetically coupled spin-1/2s. (b) An extensive
  uniform bipartition of the AKLT chain (with even-length segments)
  generates a pure antiferromagnetic spin-1/2 chain, whose ground
  state can be described by a $c = 1$ CFT. (c) An extensive random
  bipartition with even-length-only segments generates an effective
  random antiferromagnetic spin-1/2 chain, whose ground state is the
  RS$_{1/2}$ state, and (d) the corresponding $\rho_A$ in the MPS
  representation. (e) An extensive random bipartition with isolated
  spins separated by segments of even length generates an effective
  random antiferromagnetic spin-1 chain, whose ground state is the
  RS$_1$ state.}
\end{figure}

To study the quantum entanglement of the AKLT state, we can divide the chain
into A and B partitions, and define the entanglement Hamiltonian $H_{E}$
through the reduced density matrix
\begin{equation}
\mathbf{\rho }_{A}=\text{Tr}_{B}\left( |\Psi _{\mathrm{AKLT}}\rangle \langle
\Psi _{\mathrm{AKLT}}|\right) \equiv e^{-H_{E}}.
\end{equation}%
In a common practice, as illustrated in Fig.~\ref{fig:partition}a, A is an
open spin segment of length $l\geq 2$. The MPS representation dictates that $%
\rho _{A}$ contains four eigenvalues: a singlet and a triplet, as required
by the SU(2) symmetry. Explicitly, the singlet and triplet eigenvalues are~%
\cite{Fan,Katsura}
\begin{eqnarray}
\Lambda _{0} &=&\frac{1}{4}\left( 1+3\left( -\frac{1}{3}\right) ^{l}\right)
\notag \\
\Lambda _{\alpha } &=&\frac{1}{4}\left( 1-\left( -\frac{1}{3}\right)
^{l}\right) ,\quad \alpha =1,2,3.
\end{eqnarray}%
The corresponding entanglement Hamiltonian can thus be determined to be $%
H_{E}=J(l)\tau _{L}\cdot \tau _{R}$, with
\begin{equation}
J(l)=\ln \left[ \frac{1+3\left( -\frac{1}{3}\right) ^{l}}{1-\left( -\frac{1}{%
3}\right) ^{l}}\right] \simeq (-1)^{l}J_{0}e^{-l/\xi }  \label{eqn:jeff}
\end{equation}%
where we have taken the large $l$ limit, and introduced $J_{0}=4$ as the energy
unit, $\xi =1/\ln 3$ is the correlation length, $\tau _{L}$
and $\tau _{R}$ are the two fractionalized spin-1/2s at segment ends. Hence,
they are coupled antiferromagnetically for even $l$, and ferromagnetically
for odd $l$. If we swap A and B, we obtain the identical entanglement
Hamiltonian, which means that the coupling between the two end spins is
independent of their being connected by a segment in A or by that in B.

On the other hand, the bulk entanglement Hamiltonian $H_{E}$ stems from the
extensive bipartitions that divides the spin chain into A and B sets of
alternating segments (see Fig.~\ref{fig:partition}b for an example). The
word \textquotedblleft bulk\textquotedblright\ emphasizes that the
boundaries between A and B spread out the whole system, and the couplings
between the boundary spin-1/2s are relevant. Due to the gap of AKLT model,
the coupling strength between spin-1/2s decreases exponentially
with their distance, hence only nearest neighboring interactions dominate
in large $l$ limit, and the bulk entanglement Hamiltonian is described
by the spin-1/2 Heisenberg model, with coupling strength given in Eq.~(\ref{eqn:jeff}).\cite{rao16}
Hence, the $H_{E}$ with even segment length is critical and governed by
an $SU(2)_{1}$ Wess-Zumino-Witten conformal field theory (CFT) with
central charge $c=1$. This theory is, however, unstable against arbitrarily small disorder.

\section{Random bipartitioning of the AKLT chain}

A naive thinking to study the effect of quenched disorder is to randomize
the bond couplings in the parent Hamiltonian, such that we modify the ground
state wave function, hence the reduced density matrix. The AKLT case is,
however, an exception. The entanglement spectrum is immune to weak bond
randomness [i.e., if we replace uniform $J$ with random $J_{i}$ in Eq.~(\ref%
{eqn:aklt})] as the exact ground state of the random-bond AKLT model is
identical to that of the pure model.~\cite{AKLT} This is a vivid example
that the perturbation can influence thermal excitations but not the
excitations in the entanglement spectrum.

This motivates us to enforce disorder by introducing an extensive random
bipartition for the AKLT state, as illustrated in Fig.~\ref{fig:partition}c.
For the initial simplicity, we assume that the number of lattice sites in
each segment is even. After tracing out the degrees of freedom in every
other segments, the bulk entanglement Hamiltonian now reduces to a spin-1/2s
Heisenberg model with random couplings, that is%
\begin{equation}
H_{E}\simeq \sum_{i}J_{i}\tau _{i}\cdot \tau _{i+1}\text{,}  \label{Heis}
\end{equation}%
where $J_{i}=J\left( l_{i}\right) $ with $l_{i}$ being the length of the 
$i$th segment, and the form of $J\left( l\right) $ is given in Eq.~(\ref%
{eqn:jeff}). On physical ground we may then expect that the low-energy
physics is governed by the IRFP. Normally, this fixed point is revealed by a
real-space decimation process, commonly referred to as the SDRG, developed
by Dasgupta and Ma~\cite{dasgupta79} and by Fisher.~\cite{fisher94} In
general, the SDRG process in the initial stage depends strongly on the
probability distribution of the random couplings, thus often drives the
energy scale of interest to be exponentially small than the strongest bond
in the bare Hamiltonian.~\cite{wan02} Surprisingly, this is not the case
here because of the uncorrelated locations of the segment ends (which we
impose) and the finite correlation length in the topological phase.

For the distribution of segment length (hence the coupling strength), the
most natural recipe is to specify a fixed average length of the segments and
to assume that the segment ends are located independently. If we suppose
that the average length of the segment is large, one can show that the
probability distribution of the segment length $l$ (assumed to be continuous
for simplicity) satisfies the following differential equation
\begin{equation}
{\frac{dP(l)}{dl}}=-{\frac{P(l)}{\bar{l}}},
\end{equation}%
where the constant $\bar{l}$ is the average length of the segments. We can
further assume that the segments have a minimum length of $l_{0}$, such that
\begin{equation}
P(l)\sim (1/\bar{l})e^{-(l-l_{0})/\bar{l}}.  \label{eqn:poisson}
\end{equation}%
The discreteness of the segment length is not important when the average
segment length is sufficiently long. As often emerged in the minimum
mathematical model of the life expectancy problem, the exponential form of
the probability distribution simply means that the occurrence of the next
segment end is independent of the location of the previous one.

As we discussed above, the energy scale that couples two adjacent segment
end spin-1/2s depends on the length of the segment. When the segment length $%
l$ (measured by the number of sites) is sufficiently long the effective
coupling asymptotically approaches Eq.~(\ref{eqn:jeff}). For even $l$ the
effective couplings are all antiferromagnetic. We point out, though,
including ferromagnetic bonds can lead to a different SDRG fixed point with
large spins formed in a random-walk fashion.~\cite{westerberg95}

The length scales in the two previous exponential laws should in general be
different. The former is the average segment length enforced externally,
while the latter is the internal correlation length associated with the
topological phase. Together, the probability distribution of the effective
nearest-neighbor couplings reads
\begin{equation}
P(J)={\frac{1}{\Omega \Gamma }}\left( \frac{\Omega }{J}\right) ^{1-{\frac{1}{%
\Gamma }}},  \label{fpdist}
\end{equation}%
where $\Gamma ={\bar{l}/\xi }$ and $\Omega =J_{0}e^{-l_{0}/\xi }$. Once
again we stress that, in Eq.~(\ref{Heis}), the distribution of couplings $%
J_{i}$ is random and is given by Eq.~(\ref{fpdist}).

One immediately recognizes that Eq.~(\ref{fpdist}) bears a marked
resemblance to the fixed point distribution that describes the random
singlet phase in the disordered $S=1/2$ Heisenberg antiferromagnet, derived
by Fisher~\cite{fisher94,dasgupta79} by using a real space based SDRG scheme.

In Fisher's~\cite{fisher94,dasgupta79} SDRG scheme the fixed point
distribution that characterizes the random-singlet phase is obtained as the
solution of the master equation that describes the flow of the distribution
function of coupling constant:
\begin{equation}  \label{eq:master}
\begin{split}
\frac{\partial P(\beta) }{\partial \Gamma}&=\frac{\partial P(\beta) }{
\partial \beta } \\
&+P(0)\int_0^\infty d\beta_1 \int_0^\infty d\beta_2 P(\beta_1) P(\beta_2)
\delta_{\beta_1 + \beta_2 - \beta},
\end{split}%
\end{equation}
where we follow Fisher to introduce the dimensionless scaling variable $%
\beta = \ln (\Omega / J)$. The differential equation states that during the
bond decimation the flow of the bond distribution has two contributions: (i)
a shift in $\beta$ due to the reduction of the UV cutoff $\Omega$, and (ii)
the replacement of decimated bonds by the effective couplings generated
through the second-order perturbation theory as the logarithmic RG flow
parameter $\Gamma$ changes. As alluded to earlier the Eq.~(\ref{eq:master})
admits a fixed point solution of the form given in Eq.~(\ref{fpdist}).
However, at this point it is important to \emph{emphasize} that this
striking result elucidated in Eq.~(\ref{fpdist}), is not a result of any
SDRG calculation that we have performed but due to the random bipartition
scheme that we have adopted wherein the segment lengths are controlled by
the distribution given by Eq.~(\ref{eqn:poisson}). We will comment in detail
about the apparent similarity of Eq.~(\ref{eqn:poisson}), and the fixed
point distribution of the random singlet phase derived by Fisher later on in
the manuscript. At this juncture, it suffices to re-emphasize that the
extensive bipartition results in an entanglement Hamiltonian that
corresponds to an $S=1/2$ disordered Heisenberg model.

The ground state entanglement entropy of the random Hesienberg model of Eq.~(%
\ref{Heis}) was calculated by Refael and Moore~\cite{rafael04} by laying
recourse to Fisher's formulation~\cite{fisher94} of the SDRG.~\cite%
{dasgupta79} They realized that the entanglement of a segment of size $\ell $
with rest of the infinite random Heisenberg chain is just the product of the
number of singlet bonds spanning across the boundary times the entanglement
entropy per singlet (which is $\ln 2$). To accurately determine the
proportionality constant the history of the singlet bond formation across
the boundary separating the two subsystems should be kept in detail. In Ref.~%
\onlinecite{rafael04}, the SDRG methodology was very successfully adopted to
do the same, and it was shown that akin to critical systems, the
entanglement entropy in the random-singlet phase of the $S=1/2$ Heisenberg
magnet exhibits a logarithmic scaling with respect to the sub-system size $%
\ell $. In other words,
\begin{equation}
S(\ell )=\frac{\ln 2}{3}\ln \ell +\mathrm{{constant}.}  \label{eq:entropy}
\end{equation}%
Here, the coefficient $\ln 2$ that controls the logarithmic behavior is
interpreted to be an effective central charge $\tilde{c}$.

Now, in Sec.~\ref{sec:num}, we will apply the random bipartition scheme and
numerically calculate the entanglement entropy of the ground state of the
resultant entanglement Hamiltonian. In particular, we compute an effective
central charge and compare it to that of the
disordered Heisenberg chain $\tilde{c}=\ln 2$ [see Eq.~(\ref{eq:entropy})
above].



\section{Numerical Proof}

\label{sec:num}

\begin{figure}[t]
\centering
\includegraphics[width=7.5cm]{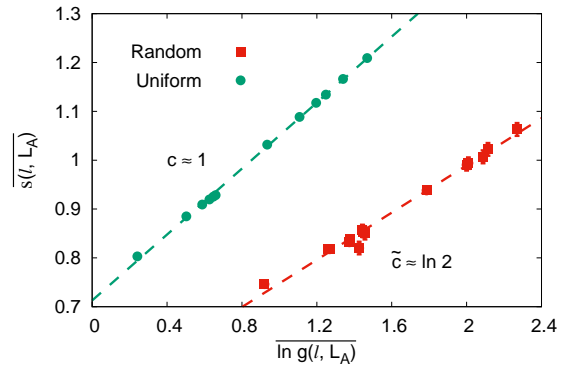}
\caption{\label{fig:nestedEE} (Color online) The ensemble-averaged
  nested entanglement entropy $\overline{s(l,L_A)}$, as a function of
  the averaged $\overline{\ln g(l,L_A)} = \overline{\ln [(L_A/\pi)
      \sin(\pi l /L_A)]}$ for the extensive random bipartition, where
  $L_A$ is the size of A, while $l$ and $L_A-l$ are the sizes of the
  nested subsystems.
The lengths of the segments are chosen according to the discrete
version of Eq.~(\ref{eqn:poisson}) with $l_0 = 2$ and $\bar{l} = 5$.
  We also plot the uniform case~\cite{rao14} for
  comparison. The slopes of the linear fits are $0.242 \pm 0.007$ and
  $0.338 \pm 0.004$, corresponding to (effective) central charges
  $\tilde{c} \approx \ln 2$ and $c \approx 1$, respectively.}
\end{figure}

To demonstrate that the IRFP is indeed accessible even in small systems, we
study the nested entanglement entropy, a straightforward generalization of
the uniform case,~\cite{rao14} and compare it to the entanglement entropy of
random $S=1/2$ Heisenberg chains, which is controlled by an effective
central charge $\tilde{c}=\ln 2$.~\cite{rafael04} For this purpose, we
impose periodic boundary condition and compute $\rho _{A}$, whose matrix
product representation is illustrated in Fig.~\ref{fig:partition}d, then determine its
ground state. We further divided the subsystem A with $n$ segments of a
total $L_{A}$ spin-1s into left partition ($p$ segments) and right partition
($n-p$ segments)
\begin{equation}
\underbrace{l_{1},l_{2},\cdots ,l_{p}}_{l=\sum_{i=1}^{p}l_{i}},\underbrace{%
l_{p+1},\cdots ,l_{n}}_{L_{A}-l}
\end{equation}%
with $l$ and $L_{A}-l$ spins, respectively. Under this nested bipartition of
A, We calculate the nested entanglement entropy $s(l,L_{A})$ for the ground
state of $\rho _{A}$. We average over random realizations of the segment
lengths in A and B, among samples with the same number of segments and the
same pattern of the nested bipartition, and plot the averaged $\overline{%
s(l,L_{A})}$ as a function of sample averaged $\overline{\ln g(l,L_{A})}=%
\overline{\ln [(L_{A}/\pi )\sin (\pi l/L_{A})]}$ in Fig.~\ref{fig:nestedEE}
for samples with 4-12 segments of random length prescribed by the discrete
version of Eq.~(\ref{eqn:poisson}) with $l_{0}=2$ and $\bar{l}=5\gg \xi $.
Depending on segment number, we choose 2,000-10,000 random realizations. For
comparison we also show data for the uniform bipartition with two spins in
each segment.~\cite{rao14} The data in the random case can be fitted by
\begin{equation}
\overline{s(l,L_{A})}={\frac{\tilde{c}}{3}}\overline{\ln \left[ {\frac{L_{A}%
}{\pi }}\sin \left( \frac{\pi l}{L_{A}}\right) \right] },
\end{equation}%
where the effective central charge $\tilde{c}=0.72\pm 0.02$ is in excellent
agreement with the expected value of $\ln 2\approx 0.693$, as oppose to $c=1$
in the uniform case.~\cite{rao14} The result is remarkable as the largest
system contains a mere 12 segments with average segment length $\bar{l}=5$
and without involving SDRG. The numerical result also demonstrates that
the exponentially small longer-range couplings among the segment end
spin-1/2s are indeed irrelevant.~\cite{fisher94}

The instant arrival at the fixed-point solution of the bulk entanglement
Hamiltonian is very reminiscent to the case of the extensive uniform
bipartition.~\cite{rao14} For the concrete example in Fig.~\ref%
{fig:partition}b, one can imagine that a spin-1/2 singlet bond is formed
within each segment, leaving an extra pair of spin-1/2s at the two ends to
be coupled with the end spins in other segments, as the result of the
bipartition. This maps the partition A (a spin-1 chain with
bipartition-induced critical dimerization) to a spin-1/2 chain with an
effective dimerization $\delta = 0$, since the lengths of A segments and B
segments are equal. Therefore, we arrive instantly at the critical point for
the pure spin-1 chain with imposed dimerization (i.e., not due to
spontaneous symmetry breaking).

Now, in the case of the random bipartition the almost instantaneous arrival
at the fixed point solution is also predicated on one another important
condition: Namely, our choice of the probability distribution for the
segment length [see Eq.~(\ref{eqn:poisson})]. It is this choice of
probability distribution that ensures that our results immediately converge
to the fixed point distribution $P(J)$ for the nearest-neighbor coupling
elucidated in Eq.~(\ref{fpdist}). Consequently, the random bipartition
effectively sets up the bond distribution that looks akin to the fixed point
distribution that characterizes the random singlet phase of the $S=1/2$
Heisenberg antiferromagnet. This provides a reason of why the numerically
computed effective central charge $\tilde{c}$ is almost bereft of finite
size effects and locks on immediately to the universal value of $\ln 2$. Of
course, this brings out immediately the question of how quickly one can
arrive at the random singlet state fixed point if we were to start with a
distribution that is non-Poissonian. We defer this discussion to the
subsequent paragraph. We end this paragraph with a few comments on the
Poissonian distribution of segment length, Eq.~(\ref{eqn:poisson}). In our
analysis we have ignored the fact that the segment length is discrete. This
can be justified if we imagine approaching the fixed point via an SDRG:
Under this SDRG, the ($\sim 1/J$) grows without limit, which means that the
system flow to infinite randomness. In our setup this points our primary
interest to the regime where $\bar{l}$ is large, hence our assumption that
the discreteness of the segment length is not important and our
simplification of the length dependence of the effective couplings $J(l)$
between the segment end spins are justified. In the same limit the choice of
$l_0$, which sets the UV cutoff, is also not important.

\begin{figure}[t]
\centering
\includegraphics[width=7.5cm]{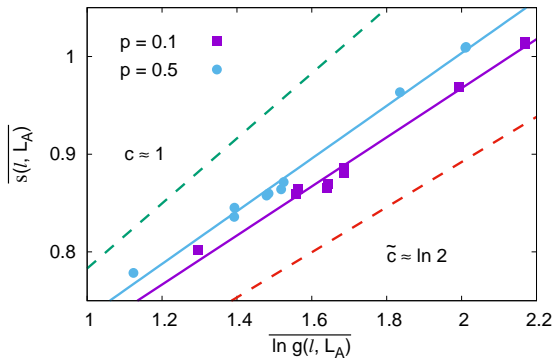} 
\caption{(Color online) The ensemble-averaged
  nested entanglement entropy $\overline{s(l,L_A)}$, as a function of
  the averaged $\overline{\ln g(l,L_A)} = \overline{\ln [(L_A/\pi)
      \sin(\pi l /L_A)]}$ for binary partition, where
  $L_A$ is the size of A, while $l$ and $L_A-l$ are the sizes of the
  nested subsystems.
  The length of each segment is chosen to be $L_1 = 4$ with probability $p$ or $L_2 = 6$
  with probability $1-p$.
The effective central charge $\tilde{c}$ can be extracted from the slope to be
$0.75 \pm 0.03$ for $p = 0.1$ and
$0.81 \pm 0.03$ for $p = 0.5$.
The values of $\tilde{c}$ are in between $c = 1$ (pure case) and
$\tilde{c} = \ln 2$ (IRFP) as indicated by the two dashed lines
(guide to eye only).
\label{fig:Binary}
}
\end{figure}

To understand how the distance to the random singlet state depends on the
segment length distribution, we consider a binary partition. The length of
each segment is chosen to be $L_1$ with probability $p$ or $L_2$ with
probability $1-p$ (with $p \leq 0.5$). Apparently, the binary distribution
is not desirable as the initial distribution for an SDRG, as the probability that 
two neighboring lengths are equal can be high. Hence, the
perturbative decimation is not valid. While it seems counterintuitive, one
can argue that the binary distribution with a smaller $p$ is closer to the
IRFP distribution than that with a larger $p$. The argument goes as follows:
With a small $p$, we expect (with high probability) to have consecutive
segments of length $L_2$ followed by a single segment of length $L_1$, then
another consecutive segments of length $L_2$ followed by a single segment of
length $L_1$, and so on. In the limit of small $p$, the number of
consecutive length-$L_2$ segments $N$ has a broad distribution (Poisson,
indeed). Within a consecutive length-$L_2$ segments (say, of total length $%
NL_2$), we have effectively $N + 1$ antiferromagnetically coupled
segment-end spin-1/2s, with low-energy excitations being magnons with energy
scale $O(J_0 e^{-L_2/\xi} /N)$. Consequently, a broader distribution of
energy scales exists naturally in the entanglement Hamiltonian for smaller $%
p $. The argument works well for $L_1 \gg L_2$, when consecutive segments of
$L_2$ are coupled weakly. Nevertheless, even for $L_1 < L_2$, the conclusion
is still valid, as the end spin-1/2s of the $L_1$ segments form singlets and
generate much weaker couplings between their neighboring segment-end spins.
In Fig.~\ref{fig:Binary}, we plot the ensemble-averaged nested entanglement
entropy $\overline{s(l,L_A)}$, as a function of the averaged $\overline{\ln
g(l,L_A)} = \overline{\ln [(L_A/\pi) \sin(\pi l /L_A)]}$ for the binary
partition with $p = 0.1$ and $0.5$. We choose $L_1 = 4 < L_2 = 6$ and find
that the effective central charge $\tilde{c} = 0.75 \pm 0.03$ for $p = 0.1$
and $\tilde{c} = 0.81 \pm 0.03$ for $p = 0.5$. We list the effective central
charge $\tilde{c}$ for five different $p$ in Table~\ref{tab:c}. As expected,
the smaller $p$ is, the closer $\tilde{c}$ is toward $\ln 2$ for the IRFP
value. On the other hand, the larger $p$ is, the closer $\tilde{c}$ is
toward 1 for the disorder-free value. We note that, as we have only up to 12
segments, finite-size artifacts can become significant for $p \leq 0.1$ so $%
\tilde{c} = \ln 2$ cannot be obtained in our calculation for binary
partition.

\begin{table}
\centering
\begin{tabular}{cccccc}
\hline\hline
\hspace{0.4cm}$p$\hspace{0.4cm} & \hspace{0.4cm}0.1\hspace{0.4cm} & \hspace{%
0.4cm}0.2\hspace{0.4cm} & \hspace{0.4cm}0.3\hspace{0.4cm} & \hspace{0.4cm}0.4%
\hspace{0.4cm} & \hspace{0.4cm}0.5\hspace{0.4cm} \\ \hline
$\tilde{c}$ & 0.75 & 0.75 & 0.76 & 0.77 & 0.81 \\
$\Delta _{c}$ & 0.03 & 0.04 & 0.03 & 0.04 & 0.03 \\ \hline\hline
\end{tabular}%
\caption{
\label{tab:c}
The effective central charge $\tilde{c}$ with uncertainty $\Delta_c$ for binary partition.
The length of each segment is chosen to be $L_1 = 4$ with probability $p$ or $L_2 = 6$ with probability $1-p$.}
\end{table}

\section{Discussions and Conclusions}

\label{sec:conclusions}

The entanglement-revealed criticality in both the pure case and the random
case establishes that exploring the appropriate bulk entanglement
Hamiltonian of the ground state wave function representing a topological
phase can efficiently distill the quantum information belonging to the
corresponding critical point separating the topological phase and its
adjacent trivial phase. This is also observed for the integer quantum Hall
transitions that can be classified by a $\mathbb{Z}$ index.~\cite{zhu14}

The distribution of the segment lengths can introduce additional relevant
perturbations for the emergent degrees of freedom. One example is that the
difference between the average segment lengths in A and B can introduce an
effective dimerization $\delta$ for the RS$_{1/2}$ state. Unlike the pure
case, $\delta$ can be \emph{continuously} tuned in the random case. The
resulting bulk entanglement Hamiltonian effectively describes random
spin-1/2 chains with weak but continuously varying bond dimerization. This
can lead to critical behavior but with finite spin correlation length
controled by $\delta$, the characteristics of a Griffiths phase.~\cite%
{hyman96} On the other hand, the random antiferromagnetic $S = 1$ chains
with enforced dimerization also contain spin-1 degrees of freedom, which,
together with the spin-1/2 degrees of freedom, result in Griffiths phases
with two independent dynamical exponents. The entanglement analogy, thus,
requires that partition A (or B) should contain segments with odd length. We
leave the details aside but point out that one can compare the extensive
random bipartition case to the domain-wall description of Damle and Huse,~%
\cite{damlehuse02} hence the relevant physics will follow.

It is interesting to point out that a spin-1 random singlet phase without
dimerization~\cite{hyman97} can also be accessed, e.g., by a bipartition
(see Fig.~\ref{fig:partition}e) that contains single sites in A (creating
effective spin-1s) and segments with random but even length in B (providing
random antiferromagnetic couplings between effective spin-1s). Indeed, it
can be shown the entanglement Hamiltonian is dominated by the
antiferromagnetic Heisenberg coupling between nearest neighboring spin-1s.
This, however, does not fit the scenario that we discussed above, because
the RS$_1$ state, as illustrated in Fig.~\ref{fig:phasediagram}, is a
critical line that is separated from the Haldane phase (including the
Griffiths region) by a multicritical point. The multicritical point has an
emergent permutation symmetry corresponding to the interchange of the
Haldane phase and the dimer phases that meet at the point.~\cite{damlehuse02}
We leave it to future investigation whether it is possible to incorporate 
the permutation symmetry and to access the multicritical point.

In summary, we have discussed how to understand the critical phase
boundaries of the antiferromagnetic Heisenberg spin-1 chain in the presence
of quenched disorder and dimerization in terms of quantum entanglement by
applying extensive bipartitions to the AKLT state that represents the
topological Haldane phase. In particular, the calculation of the effective
central charge provides a striking example how efficient quantum
entanglement can access the critical information of a topological quantum
phase transition.

\section{Acknowledgements}

This work is supported by the 973 Program under Project No. 2012CB927404 and
NSF-China through the grants No. 20121302227 and No. 11174246. X.W.
acknowledges the hospitality of the Department of Physics at the India
Institute of Technology Madras during the course of this work. RN acknowledges funding from the Visitor program 
at APCTP and also acknowledges support through NRF funded by MSIP of Korea (2015R1C1A1A01052411).


\begin{thebibliography}{99}
\bibitem{hsieh13} T. H. Hsieh and L. Fu, Phys. Rev. Lett. \textbf{113},
106801 (2014).

\bibitem{rao14} W. J. Rao, X. Wan, and G. M. Zhang, Phys. Rev. B \textbf{90}%
, 075151 (2014).

\bibitem{borchmann14} J. Borchmann \textit{et al.}, Phys. Rev. B \textbf{90}%
, 235150 (2014).

\bibitem{hsieh14} T. H. Hsieh, L. Fu, and X. L. Qi, Phys. Rev. B \textbf{90}%
, 085137 (2014).

\bibitem{santos14} R. A. Santos, J. Phys. A: Math. Theor. \textbf{48},
155203 (2015).

\bibitem{zhu14} Q. Zhu, X. Wan, and G. M. Zhang, Phys. Rev. B \textbf{90},
235134 (2014).

\bibitem{vijay14} S. Vijay and L. Fu, Phys. Rev. B \textbf{91}, 220101(R)
(2015).

\bibitem{rao16} W. J. Rao, G. M. Zhang and K. Yang, Phys. Rev. B \textbf{93}%
, 115125 (2016).

\bibitem{Li-Haldane} H. Li and F. D. M. Haldane, Phys. Rev. Lett. \textbf{101%
}, 010504 (2008).

\bibitem{Qi-2012} X. L. Qi, H. Katsura, and A. W. W. Ludwig, Phys. Rev.
Lett. \textbf{108}, 196402 (2012).

\bibitem{Chen-Wang-Lu-Lee} X. Chen, F. Wang, Y. M. Lu, and D. H. Lee, Nucl.
Phys. B \textbf{873}, 248 (2013).

\bibitem{Haldane-1983} F. D. M. Haldane, Phys. Lett. \textbf{93A}, 464
(1983); Phys. Rev. Lett. \textbf{50}, 1153 (1983).

\bibitem{AKLT} I. Affleck, T. Kennedy, E. H. Lieb, and H. Tasaki, Phys. Rev.
Lett. \textbf{59}, 799 (1987); Commun. Math. Phys. \textbf{115}, 477 (1988).

\bibitem{Pollmann-Oshikawa-2010} F. Pollmann, A. M. Turner, E. Berg, and M.
Oshikawa, Phys. Rev. B \textbf{81}, 064439 (2010).

\bibitem{huse01} See references in D. Huse, Phys. Rep. \textbf{348}, 159
(2001).

\bibitem{dasgupta79} C. Dasgupta and S. K. Ma, Phys. Rev. B \textbf{22},
1305 (1980).

\bibitem{fisher94} D. S. Fisher, Phys. Rev. B \textbf{50}, 3799 (1994).

\bibitem{hyman97} R. A. Hyman and K. Yang, Phys. Rev. Lett. \textbf{78},
1783 (1997).

\bibitem{monthus97} C. Monthus, O. Gollineli, and Th. Jolicoeur, Phys. Rev.
Lett. \textbf{79}, 3254 (1997); Phys. Rev. B \textbf{58}, 805 (1998).

\bibitem{affleck87} I. Affleck and F. D. M. Haldane, Phys. Rev. B \textbf{36}%
, 5291 (1987).

\bibitem{damle02} K. Damle, Phys. Rev. B \textbf{66}, 104425 (2002).

\bibitem{Fan} H. Fan, V. Korepin, and V. Roychowdhury, Phys. Rev. Lett.
\textbf{93}, 227203 (2004).

\bibitem{Katsura} H. Katsura, T. Hirano, and Y. Hatsugai, Phys. Rev. B
\textbf{76}, 012401 (2007).

\bibitem{wan02} X. Wan, K. Yang, and R. N. Bhatt, Phys. Rev. B \textbf{66},
014429 (2002).

\bibitem{westerberg95} E. Westerberg, A. Furusaki, M. Sigrist, and P. A.
Lee, Phys. Rev. Lett. \textbf{75}, 4302 (1995).

\bibitem{rafael04} G. Refael and J. E. Moore, Phys. Rev. Lett. \textbf{93},
260602 (2004).

\bibitem{laflorencie} Nicolas Laflorencie, Phys. Rev. B\textbf{72},
140408(R), 2005.

\bibitem{hyman96} R. A. Hyman, K. Yang, R. N. Bhatt, and S. M. Girvin, Phys.
Rev. Lett. \textbf{76}, 839 (1996).

\bibitem{damlehuse02} K. Damle and D. A. Huse, Phys. Rev. Lett. \textbf{89},
277203 (2002).

\end{thebibliography}
\end{document}